# INTEGRAL: status of the mission – after 10 years

**Christoph Winkler[1]**
*Research and Science Support Department, Astrophysics and Fundamental Missions Division*
*ESA-ESTEC, Keplerlaan 1, NL-2201 AZ Noordwijk, The Netherlands*
*E-mail:* `cwinkler@rssd.esa.int`

The ESA gamma-ray observatory INTEGRAL, launched on 17 October 2002, continues to produce a wealth of discoveries and new results on compact high energy Galactic objects, nuclear gamma-ray line emission, diffuse line and continuum emission, cosmic background radiation, AGN, high energy transients and sky surveys. Ten years after launch, the spacecraft, ground segment and payload are in excellent state-of-health, and INTEGRAL is continuing its scientific operations well beyond its 5-year technical design lifetime until, at least, 31 December 2014. This paper summarizes the current status of INTEGRAL.

---

[1] Speaker

# 1. The INTEGRAL mission

## 1.1 Overview

The ESA observatory INTEGRAL[2] [1] is dedicated to the fine spectroscopy (2.5 keV FWHM @ 1 MeV) and fine imaging (angular resolution: 12 arcmin FWHM) of celestial gamma-ray sources in the energy range 15 keV to 10 MeV with concurrent source monitoring in the X-ray (3-35 keV) and optical (V-band, 550 nm) bands. INTEGRAL, with a total launch mass of about 4 tons, was launched on 17 October 2002 from the Baikonur Cosmodrome (Kazakhstan) using a PROTON rocket equipped with a Block DM $4^{th}$ stage. The orbit is characterized by a high perigee in order to provide long periods of uninterrupted observations with nearly constant background and away from trapped radiation (electron and proton radiation belts). The orbital parameters at the beginning of the mission were: 72-hour orbit with an inclination of 52.2 degrees, a height of perigee of 9000 km and a height of apogee of 154000 km.

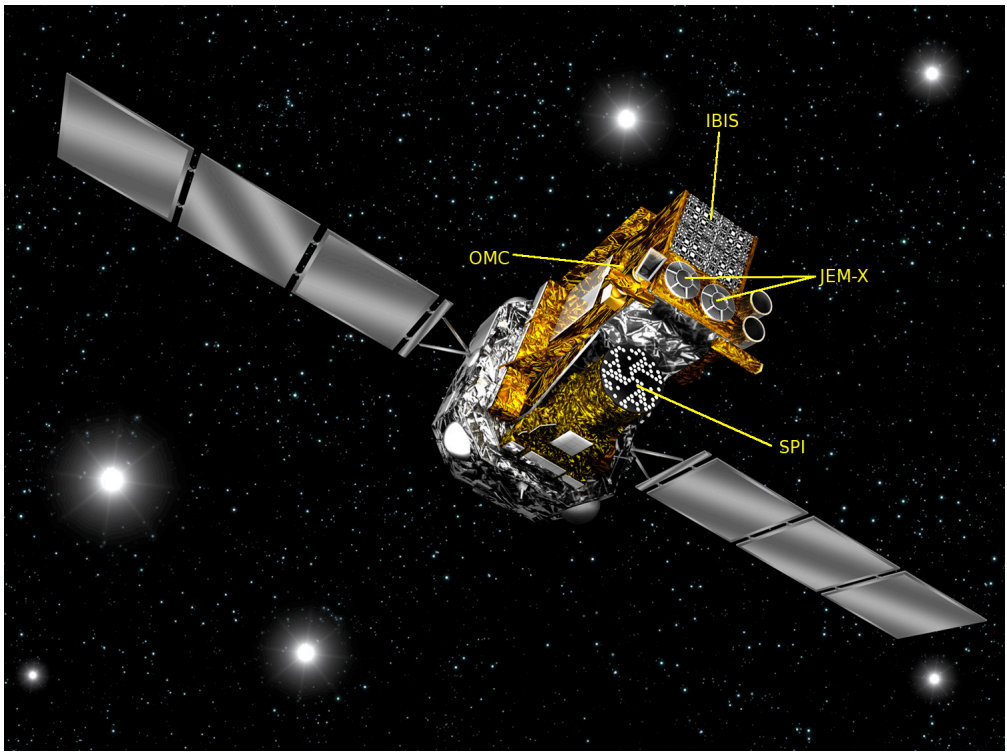

***Figure 1:*** *The INTEGRAL spacecraft. Courtesy ESA (C. Carreau).*

Owing to background radiation effects in the high-energy detectors, scientific observations are carried out while the satellite is above a nominal altitude of typically 40000 to 60000 km (i.e. above the van Allen belts). This means, that most of the time spent in the orbit provided by the PROTON launcher can be used for scientific observations, about 210 ks per revolution. An on-board particle radiation monitor allows the continuous assessment of the radiation environment

---

[2] <u>INT</u>ERNATIONAL <u>G</u>AMMA-<u>R</u>AY <u>A</u>STROPHYSICS <u>L</u>ABORATORY

local to the spacecraft. INTEGRAL carries two main gamma-ray instruments, the spectrometer SPI [2], optimized for the high-resolution (2.5 keV FWHM @ 1 MeV) gamma-ray line spectroscopy (20 keV - 8 MeV), and the imager IBIS [3], optimized for high angular resolution (12 arcmin FWHM) imaging (15 keV - 10 MeV). Two monitors, JEM-X [4] in the (3-35) keV X-ray band, and OMC [5] in the optical Johnson V-band complement the payload. All instruments are co-aligned with overlapping fully coded field-of-views ranging from 4.8º diameter (JEM-X), 5º×5º (OMC), to 9º×9º (IBIS) and 16º corner-to-corner (SPI), and they are operated simultaneously, providing the observer with data from all four instruments. Besides high spectral and high angular resolution, INTEGRAL also offers ms timing resolution and polarimetry capabilities. The Mission Operations Centre at ESOC (Darmstadt/Germany) performs all standard spacecraft and payload operations and maintenance tasks. The INTEGRAL Science Operations Centre (ISOC) [6] in Madrid (Spain) is responsible for the science operations planning including the implementation of Target of Opportunity observations within the pre-planned observing programme. The INTEGRAL Science Data Centre (ISDC) [7] in Versoix (Switzerland) receives the science telemetry for near-real time monitoring, standard science analysis and archiving.

## 1.2 Current programmatic and technical status

The nominal 2-year mission operations phase was completed on 1 January 2005, and INTEGRAL is currently being operated in the extended mission phase. Several mission extension requests have been granted since then, and INTEGRAL mission operations are currently funded until 31 December 2014. A new request to extend the mission by two more years, until 31 December 2016, was made in Fall 2012 and is currently under review by ESA. Ten years after launch, the INTEGRAL spacecraft continues to operate flawlessly! The payload is in a very good shape: the percentage of healthy detectors is 79% (SPI), 96% (IBIS-ISGRI), 98% (IBIS-PICsIT), 76% (JEM-X), and 100% (OMC). As of 10 October 2010 (revolution 976), both JEM-X units are being operated simultaneously again.

During the 10 years since launch, 70 kg of on-board fuel have been used for attitude control and orbit maintenance: an average consumption of about 0.6 kg/month. However, 110 kg of fuel are still available. Likewise, the solar array power margin is very comfortable: the spacecraft is still being operated in its nominal configuration, which thanks to its solar aspect angle constraint of ±40°, allows a good sky visibility at any point in time, until at least 2016. Only then, the steady degradation of the solar arrays is expected to reduce the array's power output such that the solar aspect angle will have to be constrained to ±30°.

Solar and lunar gravitation are influencing the orbital parameters. For example, the perigee height evolved from 9000 km at the start of the mission to about 13000 km in 2007 and has reached a minimum of 2756 km on 25 October 2011, after which it will increase again to a local maximum of 10000 km in 2016 with subsequent decrease. As of early 2010, the solar activity is picking up towards the next solar maximum in 2013/2014. The corresponding cosmic-ray induced background, being anti-correlated with the solar activity, shows a decreasing background rate in all detectors and veto subsystems since early 2010.

## 2. The observing programme

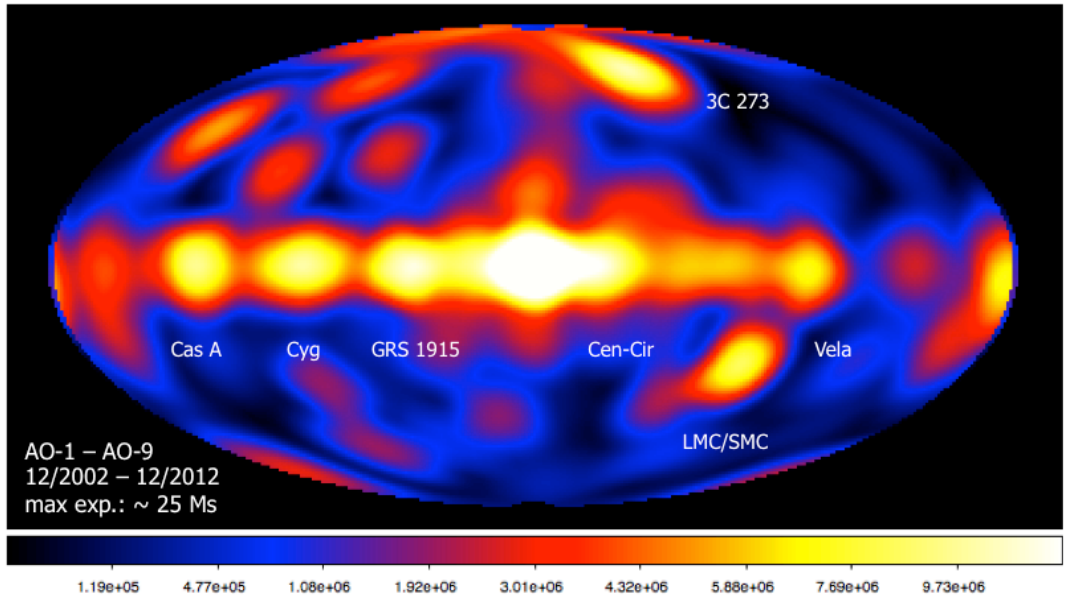

*Figure 2:* *The total exposure map for observations from Dec 2002 until Dec 2012 (AO-1 until AO-9) in galactic co-ordinates. The color-scale units are seconds. Courtesy ESA/ISOC (G. Bélanger).*

INTEGRAL is being operated as an observatory. The observing programme, 100% open (as of 2009) to the scientific community at large is built from observing proposals, which have been peer reviewed by the Time Allocation Committee (TAC). Submission of observing proposals is invited following an Announcement of Opportunity (AO), issued by ESA once a year. In the early years of the mission, the annual observing programme also contained a guaranteed programme ("Core Programme" [8]), which used between 20% and 35% of the observing time per year, and which provided science return to the instrument and ISDC PI teams and to other members of the Science Working Team. In 2007 (AO-4), ESA introduced a long (~ Ms) observation of the Galactic Centre as a pilot "key programme" to the community. The concept of key (legacy) programmes turned out to be very successful and tailor-made for INTEGRAL: a key programme should be at least as long as 1 Ms, and can span two years (two AO-cycles) if justified by its science objectives. The vast majority of approved open time observations during the past years consist of these key programmes. All observations, with the exception of ToO follow-up observations, are open to the scientific community for the submission of so-called "data rights proposals": successful proposers are granted data rights on those sources which are contained in the very large FOV (~ 100 square degrees, fully coded) of accepted observations, but which had not been requested by the PIs of those observations. The routine scheduling of accepted open time observations during a calendar year shows that the typical duration of observations ranges between 500 ks and 3 Ms. Shorter exposures (< 500 ks) are performed mostly for ToO follow-up observations. Pointed observations are usually being performed using the standard 5×5 point or the 7-point hexagonal dither pattern centered on the main target, but about 20% of the annual observing time is, on average, spent on observations using scans or customized dither patterns. About one third of all accepted ToO observations are based on unsolicited notifications of new and unforeseen ToO events, i.e. outside the TAC review process of AO-based ToO proposals.

With the AO-9 cycle of observations being completed on 31 December 2012, the total science observing time, which has been used with INTEGRAL over the time period from 30 December 2002 until 31 December 2012 amounts to about 245 Ms. The total exposure map over this period is shown in Figure 2.

## 3. Science community interfaces

The science community can interface with INTEGRAL through a number of channels: Scientists can access public and private[3] data from various archives: (1) The ISDC is providing consolidated and near-real time data. (2) Archival data are available at the ISDC; at the ISOC; at the INTEGRAL Guest Observer Facility at NASA/GSFC; and at the Russian Science Data Centre at IKI (Moscow). (3) Instantaneous public data are available from: (i) selected AO observations on PI request (e.g., galactic bulge monitoring and galactic plane scan programmes); (ii) in-flight calibration observations; (iii) Earth occultation/CXB observations; (iv) nearby supernovae (SN Ia < 1 Mpc, SN II < 60 kpc); (v) ToO follow-up observations with wide general interest (e.g. SN 2011fe, Crab flares, AXP 1E1547.0-5408, Swift J174510.8-262411); (vi) GRB in the FOV (location, light curve, peak flux, fluence, duration); (vii) GRB outside the FOV (SPI/ACS light curves). Target of Opportunity (ToO) observations provide important very serendipitous science. These observations are possible via the execution of TAC approved ToO proposals, which will be scheduled once their specific scientific and operational criteria have been fulfilled. However, in case of a new (unexpected) event, scientists can always alert the INTEGRAL science operations team in order to consider an observation, which is timely and of scientific relevance, even if it is not included in the database of accepted observing programmes of the on-going AO cycle. The scientific users community at large can actively interface with the INTEGRAL project by joining two important international committees[4], with a membership for a few years on a rotational basis: (i) The INTEGRAL Time Allocation Committee, which is in charge of peer reviewing all observing and data rights proposals. (ii) The INTEGRAL Users Group, acting as a focus for the interests of the scientific community in INTEGRAL, and as an advocate for INTEGRAL within that community.

## 4. Science Highlights

Since its launch in 2002, INTEGRAL has produced until October 2012 about 2000 publications (including about 700 papers in refereed journals), and at least 90 PhD theses have been completed using data and results of the mission. The summary below contains selected scientific highlights, some of which have been addressed in more detail during this workshop.

---

[3] Private (PI) data will become public after one year.
[4] See for more information: http://www.rssd.esa.int/index.php?project=INTEGRAL&page=Teams

*Nucleosynthesis and gamma-ray lines*
- The first large-scale sky-map at 511 keV
- $^{44}$Ti lines fom SNRs Cas A and 1987A
- The Galaxy-wide origin of the $^{26}$Al gamma-ray line
- Determination of the Galactic cc-SN rate
- The $^{26}$Al/$^{60}$Fe yield ratio constraining SN models

*Galactic compact objects*
- A new class of X-ray binaries enshrouded by massive stellar wind ("strongly absorbed HMXB")
- A new regime of intermittent accretion in wind-fed relativistic binaries ("SFXT")
- Polarized hard X-ray emission in Crab, Cyg X-1 and in GRB 041219a
- Dominance of accreting WD in the "diffuse" galactic ridge hard X-ray emission
- Hard spectral tails in extremely magnetized neutron stars: $e^{\pm}$ pairs dominated fireball
- Strong soft γ-ray emission during magnetar (AXP) outbursts
- Cyclotron line strength correlated with the height of the NS accretion column
- Pulsed hard X-ray emission up to ~150 keV from rotation-powered pulsars
- The past activity of Sgr A* monitored via its Compton echo
- The Crab nebula is not a stable in-flight calibration source
- Diffuse hard X-ray emission from the Vela pulsar wind nebula

*Extragalactic objects*
- The first measurement of the true fraction of Compton-thick AGN
- The hard X-ray luminosity function of AGN
- The most distant QSO seen in hard X-rays
- A population of faint GRB with long spectral lags